\documentclass[prb,twocolumn,amsmath,amssymb,superscriptaddress,showpacs]{revtex4}

\usepackage{graphicx}

\def\eps{\varepsilon}

\begin{document}

\title{Hybridization of wave functions in one-dimensional localization}

\author{D.~A.~Ivanov}
\affiliation{Institute of Theoretical Physics,
Ecole Polytechnique F\'ed\'erale de Lausanne (EPFL),
CH-1015 Lausanne, Switzerland}

\author{M.~A.~Skvortsov}
\affiliation{L.~D.~Landau Institute for Theoretical Physics, 142432 Chernogolovka,
Russia}
\affiliation{Moscow Institute of Physics and Technology, 141700 Dolgoprudny, Russia}

\author{P.~M.~Ostrovsky}
\affiliation{Institut f\"ur Nanotechnologie, Forschungszentrum Karlsruhe,
76021 Karlsruhe, Germany}
\affiliation{L.~D.~Landau Institute for Theoretical Physics, 142432 Chernogolovka,
Russia}

\author{Ya.~V.~Fominov}
\affiliation{L.~D.~Landau Institute for Theoretical Physics, 142432 Chernogolovka,
Russia}
\affiliation{Moscow Institute of Physics and Technology, 141700 Dolgoprudny, Russia}

\begin{abstract}

A quantum particle can be localized in a disordered potential,
the effect known as Anderson localization. In such a system,
correlations of wave functions at very close energies may be described,
due to Mott,
in terms of a hybridization of localized states. We revisit
this hybridization description and show that it may be used to obtain
quantitatively exact expressions for some asymptotic features
of correlation functions, if the tails of the wave functions
and the hybridization matrix elements are assumed to have
log-normal distributions typical for localization effects.
Specifically, we consider three types of one-dimensional
systems: a strictly one-dimensional wire and two
quasi-one-dimensional wires with unitary and orthogonal
symmetries. In each of these models, we consider two
types of correlation functions: the correlations of the density
of states at close energies and the dynamic response
function at low frequencies. For each of those correlation
functions, within our method, we calculate three asymptotic
features: the behavior at the logarithmically large
``Mott length scale'', the low-frequency
limit at length scale between the localization length and
the Mott length scale, and the leading correction in frequency
to this limit. In the several cases, where exact results are
available, our method reproduces them
within the precision of the orders in frequency considered.
\end{abstract}

\date{December 22, 2011}


\maketitle

\section{Introduction}
\label{sec:intro}

The localization of a quantum particle in a disordered potential (commonly
known as Anderson localization \cite{Anderson1958}) is one of the most
fascinating mesoscopic phenomena (see, e.g.,
Refs.\ \onlinecite{LeeRamakrishnan1985} and \onlinecite{KramerMackinnon1993}
for a review). Arising from quantum interference between different particle
trajectories, localization depends strongly on the dimensionality of the
system. In one dimension, such an interference is most relevant, and an
arbitrarily weak potential is known to localize a particle (in the absence
of decoherence) \cite{MottTwose61,GertsenshteinVasilev1959,Thouless77,Dorokhov}.
Besides, one-dimensional case is most accessible for
analytic studies, which makes it the best understood model of
localization (see, e.g., Refs.\ \onlinecite{Beenakker1997} and
\onlinecite{Mirlin-review}).

For the purpose of the present paper, we distinguish several models of
one-dimensional localization: the {\em strictly-one-dimensional} (S1D) case
(with one conducting channel) and the {\em quasi-one-dimensional} (Q1D) wire
(with $N\gg 1$ conducting channels). These two limits exhibit some common
universal properties, but are typically treated with different analytic
techniques (Berezinsky technique \cite{Berezinsky73,BerezinskyGorkov79}
and contemporary methods\cite{OssipovKravtsov06,KravtsovYudson11}
in the S1D case,
and the sigma-model technique \cite{Efetov83,Efetov-book,EL83} in the Q1D case).
The quasi-one-dimensional wires may be further classified in terms
of the symmetries of the Hamiltonian, according to the random-matrix-theory
scheme (unitary, orthogonal, etc.) \cite{Efetov-book,Mehta}.

One of the main quantitative characteristics of the
localization is the statistics of (localized) eigenfunctions.
In one dimension, it was studied extensively, and many
analytic results are available
\cite{Mirlin-review,Mirlin-JMP-1997,Gogolin,LGP82,Kolokolov1995}.
Most of the analytical
results are derived in the weak-disorder regime (which is
believed to obey the single-parameter-scaling
property \cite{AALR79,ATAF80,Shapiro86}, see also
Refs.\ \onlinecite{KRS88}, \onlinecite{DLA00}, and \onlinecite{Beenakker1997}
for further discussions). Remarkably, the statistics
of the ``envelopes'' of localized eigenfunctions in this
regime is universal for S1D and Q1D problems
(independently of the symmetry class) and can be expressed in terms
of the Liouville quantum mechanics \cite{Mirlin-review,Mirlin-JMP-1997},
while the short-range oscillations
distinguish between S1D and Q1D cases and between symmetry classes in the
Q1D case.

\begin{figure}[tb]
\centerline{\includegraphics[width=0.32\textwidth]{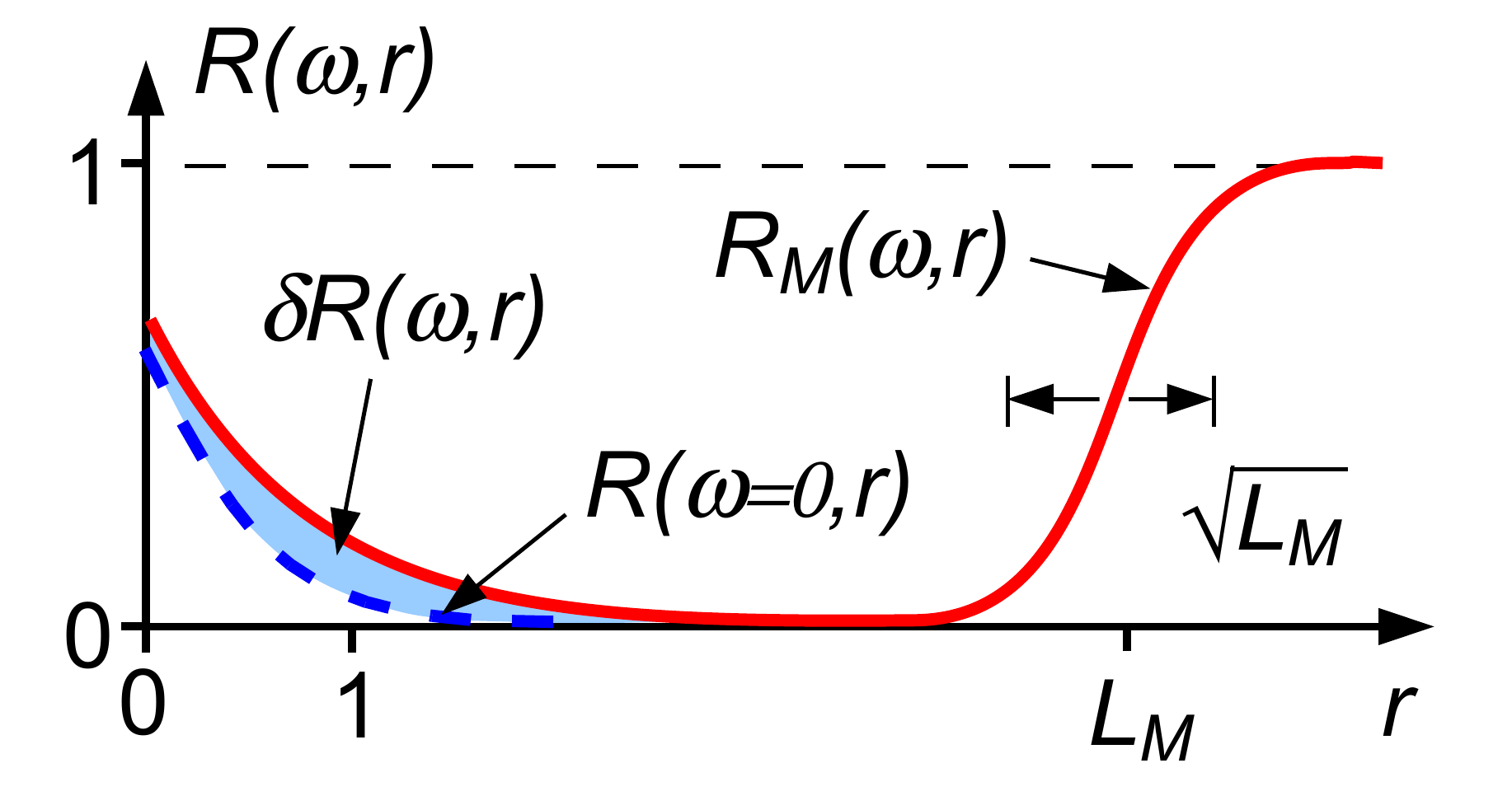}}
\centerline{\includegraphics[width=0.32\textwidth]{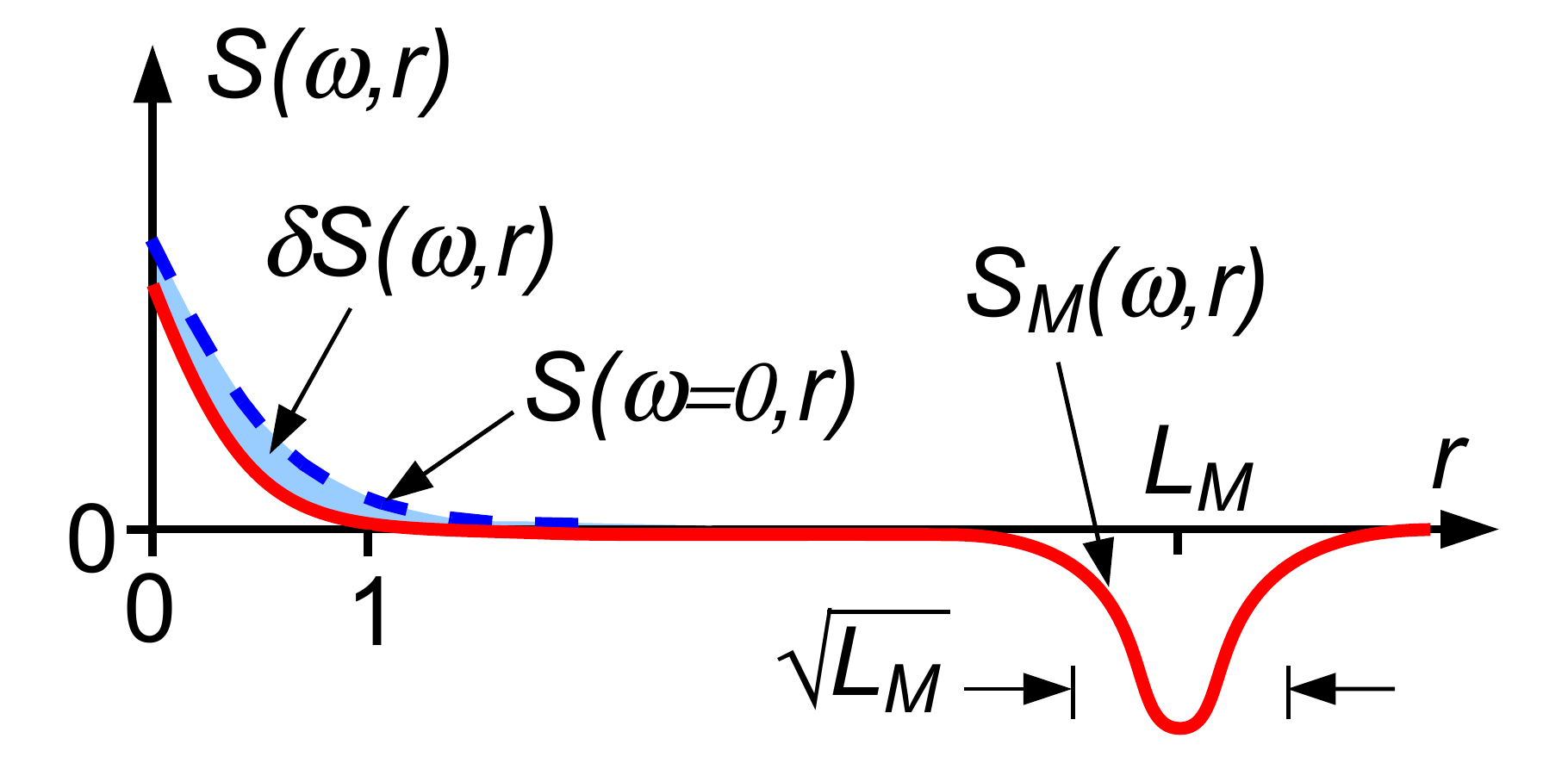}}
\caption{A schematic (not numerically exact)
view of the correlation functions $R(\omega,r)$
(top panel) and $S(\omega,r)$ (bottom panel)
defined in Eqs.~(\ref{R-function}) and (\ref{S-function}), respectively.
The dashed lines denote
the $\omega\to 0$ limits [the same function (\ref{single-function-exact})
for  $R(\omega,r)$ and $S(\omega,r)$]. The shaded regions denote
$\delta R(\omega,r)$ and $\delta S(\omega,r)$ as defined in
Eq.~(\ref{deltaRS}). The features at the Mott length scale $L_M$
are denoted by $R_M(\omega,r)$ and $S_M(\omega,r)$, respectively.
}
\label{fig:RS-plots}
\end{figure}

\begin{table*}[ht!]
\caption{Correlation function $R(\omega,r)$ at low frequencies ($\omega\ll1$).}
\label{table:R}
\begin{tabular}{|c|c|c|c|}
\hline
Model & $R(\omega\to 0,r\gg 1)$ & $\delta R(\omega,r\gg 1)$ & $R_M(\omega,r)$ \\
\hline
S1D   & & $\propto \omega^2(L_M - 3r) e^{2r}$ & \\
\cline{1-1} \cline{3-3}
Q1D-unitary & $\propto r^{-3/2} e^{-r/4}$ &
$\propto \omega^2(L_M - 3r)^2 e^{2r}$ &
\raisebox{0pt}[0pt][0pt]{$\displaystyle \frac{1}{2} \left(1+{\rm erf} \frac{r-L_M}{2\sqrt{r}} \right)$}  \\
\cline{1-1} \cline{3-3}
Q1D-orthogonal & & $\propto \omega e^{r/2}$ & \\
\hline
\end{tabular}
\end{table*}

\begin{table*}[ht!]
\caption{Correlation function $S(\omega,r)$ at low frequencies ($\omega\ll1$).}
\label{table:S}
\begin{tabular}{|c|c|c|c|}
\hline
Model & $S(\omega\to 0,r\gg 1)$ & $\delta S(\omega,r\gg 1)$ & $S_M(\omega,r)$ \\
\hline
S1D   & & $\propto - \omega^2(L_M - 3r) e^{2r}$ & \\
\cline{1-1} \cline{3-3}
Q1D-unitary & $\propto r^{-3/2} e^{-r/4}$ &
$\propto - \omega^2(L_M - 3r)^2 e^{2r}$ &
\raisebox{0pt}[0pt][0pt]{$\displaystyle  - \frac{1}{2\sqrt{\pi r}}
\exp \left[-\frac{(r-L_M)^2}{4 r} \right]$} \\
\cline{1-1} \cline{3-3}
Q1D-orthogonal & & $\propto - \omega e^{r/2}$ & \\
\hline
\end{tabular}
\end{table*}

A more detailed information about localization (in particular, relevant to
dynamic properties) can be extracted from correlations between
eigenfunctions at different energies. Two such quantities may be defined
\cite{GDP1,GDP2}:
the density-of-states (DOS) correlation function,
\begin{multline}
  R(\omega, |x_1-x_2|)
  =
  \nu^{-2}\,
  \Bigl<
  \sum_{n,m} \delta(E_n-E) \delta(E_m-E-\omega)
\\ {}
  \times |\psi_n(x_1)|^2 |\psi_m(x_2)|^2
  \Bigr> \, ,
\label{R-function}
\end{multline}
and the dynamic response function,
\begin{multline}
  S(\omega, |x_1-x_2|)
  =
  \nu^{-2}\,
  \Bigl<
  \sum_{n,m} \delta(E_n-E) \delta(E_m-E-\omega)
\\ {}
  \times \psi^*_n(x_1) \psi_n(x_2) \psi^*_m(x_2) \psi_m(x_1)
  \Bigr> \, .
\label{S-function}
\end{multline}
Here the sum is taken over the eigenstates $\psi_n$ with energies $E_n$, and
$\nu$ is the average density of states. The normalization of these correlation
functions is chosen in such a way that they are dimensionless quantities with
a finite limit in an infinitely long wire.
It will be furthermore convenient to measure the lengths in the units of the
localization length $\xi$ and the energies in the units of the average
level spacing within the localization length
$\Delta_\xi$ \cite{loc-length-definition}.
With this convention, $R(\omega, r)$ and $S(\omega, r)$
become dimensionless functions of dimensionless parameters.

Both $R(\omega, r)$ and $S(\omega, r)$ were studied analytically in detail
in the S1D model \cite{GDP1,GDP2}, and $R(\omega,r)$
has been recently calculated in the Q1D-unitary model \cite{IOS2009}
(in all those studies, the weak-disorder limit
was assumed). While the limiting form of these correlations at $\omega \to 0$
is determined by the single-wave-function statistics and is therefore
universal for both S1D and Q1D models, the corrections at finite $\omega$
distinguish between S1D and Q1D \cite{SO07,IOS2009}.
Qualitatively, the properties of
the correlation functions $R(\omega, r)$ and $S(\omega, r)$
in the low-frequency limit $\omega \ll 1$
may be understood
using the original argument by Mott about the hybridization of the localized
wave functions \cite{Mott}
(the opposite limit $\omega \gg 1$ can be studied by means
of the standard perturbation theory).
However, the first attempt to promote
the Mott's arguments to quantitative calculations in
Ref.\ \onlinecite{SivanImry1987} produced some
inaccurate results (as can be seen by comparing to exact
expressions \cite{GDP1}), since it neglected mesoscopic fluctuations of the
tails of the localized states.

In the present work, we rectify this approach and revisit
Mott's arguments on wave-function hybridization \cite{Mott} taking into account
the log-normal distribution of the tails of the localized
states \cite{Mirlin-review}.
Our method is based on a number of assumptions that we introduce
in the main text and then explicitly summarize and discuss in
the last section of the paper (Section~\ref{sec:summary}).
As a result,
we obtain a semi-phenomenological description of the hybridization of the
localized states at distances much larger than the localization length,
$r \gg 1$. Our theory reproduces correctly the physics at the ``Mott
length scale'',
\begin{equation}
L_M = 2 \ln (1 / \omega) ,
\label{Mott-length}
\end{equation}
and the leading correction to $R(\omega, r)$ at $1 \ll r < L_M$ in the
Q1D-unitary model. We further make predictions concerning the properties
of $R(\omega, r)$ in the Q1D-orthogonal model and of $S(\omega,r)$ in
all the above-mentioned models. These predictions may be checked against
future sigma-model calculations in Q1D systems.

\section{Main results}
\label{sec:results}

In the present work, we assume the single-parameter-scaling regime
for the tails of the localized states. Namely, we suppose that
at distances $r\gg 1$, the decay of the localized wave function
may be described by a log-normal distribution
with the width and median
(or, formally, the variance and the mean of the
logarithm)
described by one parameter and with an appropriate
cut-off of the tails. By combining this assumption with the Mott's argument
about the wave-function hybridization (see subsequent sections for details),
we can infer quantitative details about the behavior of the correlation
functions (\ref{R-function}) and (\ref{S-function})
in the low-frequency limit $\omega \ll 1$.

The general structure of those two correlation functions contains two
main separate regimes: $r \ll L_M$ and $r \sim L_M$ (Fig.~\ref{fig:RS-plots}).
At $r \ll L_M$, the correlations are known to be dominated by the
statistics of a single wave function \cite{Mott,SivanImry1987,GDP1,GDP2,IOS2009},
and it is natural to represent them as
\begin{subequations}
\label{deltaRS}
\begin{gather}
R(\omega,r) = R(\omega\to0,r) + \delta R(\omega,r)\, ,
\\
S(\omega,r) = S(\omega\to0,r) + \delta S(\omega,r)\, ,
\end{gather}
\end{subequations}
where $\delta R(\omega,r)$ and $\delta S(\omega,r)$ vanish as $\omega\to 0$.

At $r\sim L_M$, the correlation function $R(\omega,r)$ exhibits a crossover
from zero to one centered at $L_M$ and with a width of the order $\sqrt{L_M}$,
and the correlation function $S(\omega,r)$ has a negative
bump at the same location \cite{GDP1,GDP2}.
The asymptotic form of these features at
$\omega \to 0$ will be denoted as $R_M(\omega,r)$ and  $S_M(\omega,r)$,
respectively.

Our hybridization argument reproduces the (universal) main asymptotics
$R(\omega\to0,r)$ and $S(\omega\to0,r)$
at $ 1\ll r \ll L_M$, the (nonuniversal)
leading in $\omega$ corrections $\delta R(\omega,r)$ and $\delta S(\omega,r)$,
as well as the universal
behavior of $R_M(\omega,r)$ and  $S_M(\omega,r)$, see Tables \ref{table:R}
and \ref{table:S} \cite{loc-length-warning}. Some of these results can be
verified against the existing exact calculations, while others present
new conjectures. As a byproduct of our calculation, we also relate the
$R(\omega\to0,r)$ and $S(\omega\to0,r)$ to the statistics of a single
wave function [Eq.~(\ref{two-vs-single-1d}) below], including
the proportionality coefficient.

\section{Statistics of wave-function tails}
\label{sec:tails}

We start with a simplified statistical description of a single
localized state in terms of the log-normal distribution of its
tails. The statistics of a single wave function has been studied
in detail in both Q1D and S1D geometries \cite{Mirlin-review,Kolokolov1995},
and we first briefly summarize the existing results and then propose
our approximation.

First of all, a localized state $\psi(x)$ can be represented as a product
of a slowly varying envelope and a rapidly oscillating short-range
component \cite{Mirlin-review}:
\begin{equation}
\psi(x)=\widetilde\psi(x) \cdot \varphi(x) \cdot (A\xi)^{-1/2}
\label{wf-decomposition}
\end{equation}
(here we include the dimensional factor $(A\xi)^{-1/2}$, where
$A$ is the cross section of the wire, in order to simplify the formulas
below).
The short-range component $\varphi(x)$ is correlated on the scale of
the mean free path $l$ and oscillates on the scale of the particle
wave length $\lambda_F$. We choose it normalized to
$\langle |\varphi(x)|^2\rangle =1$. The ``envelope'' component
$\widetilde\psi(x)$ is correlated on the scale of the localization
length $\xi$ (in S1D, $\xi \sim l$; in Q1D, $\xi \gg l$) and does
not oscillate. The two components $\widetilde\psi(x)$ and
$\varphi(x)$ are distributed independently.
It was shown in Refs.~\onlinecite{Mirlin-review,Kolokolov1995}
that such a decomposition
is exact with the statistics of $\widetilde\psi(x)$ being
universal for Q1D and S1D systems, and that of
$\varphi(x)$ distinguishing between S1D and Q1D and between
different symmetry classes in Q1D.

The statistics of $\widetilde\psi(x)$ can be most conveniently
described in terms of its logarithm
\begin{equation}
\chi(x)=\ln |\widetilde\psi(x)|^2\, .
\label{chi-def}
\end{equation}
As shown in Ref.~\onlinecite{Mirlin-review} (section 3.2.2), the
statistics of $\chi(x)$ is given by a functional integral,
which involves a diffusion-type quadratic action in $\chi(x)$
and a delta-function
constraint imposing the normalization of the wave function
$\int e^\chi dx=1$.

An accurate treatment of that functional integral is difficult, and
we simplify it by observing that the tails of $\psi(x)$ contribute
very little to the normalization, and therefore the normalization-enforcing
delta-function term is of little importance for the tails of $\psi(x)$.
The normalization is mostly determined by the maxima of $\chi(x)$, and
therefore the main role of this delta-term is to normalize
the maxima of the function $\chi(x)$. We expect that the distribution of the
maxima of $\chi(x)$ has a width of order one and centered around zero.

This suggests our approximation for studying the wave-function tails.
Instead of working with a full path integral of
Ref.~\onlinecite{Mirlin-review}, we first fix the position $x_0$ and
the value $\chi(x_0)$ of the maximum of $\chi(x)$
(with $|\chi(x_0)| \lesssim 1$) and replace the normalization constraint
by an approximate condition that $\chi(0)<\chi(x_0)$ everywhere. This
guarantees a normalization of the wave function to a ``logarithmic''
precision: namely, the normalization of wave functions constructed in
such a way will be of order one. Furthermore, we will only be interested
in a ``coarse-grained'' behavior of $\chi(x)$: typical scales of
interest of $\chi(x)$ will be of order $r$, and therefore for many
purposes we do not need to distinguish between the maximum value
$\chi(x_0)$ and zero.

We thus arrive at the following coarse-grained description of
the ensemble of the envelopes $\chi(x)$: a localized state
is determined by the location $x_0$ of its (global) maximum (where
$\chi(x_0)\approx 0$) and the functional measure
for the tails (which follows directly
from the formula (3.34) of
Ref.~\onlinecite{Mirlin-review})
\begin{multline}
d\mu_{x_0} [\chi(x)] \propto \\
\exp \left(-\int \frac{1}{4} \left[ \frac{d\chi}{dx} +
{\rm sign}(x-x_0)\right]^2 dx \right)\, D\chi(x)
\label{tail-action}
\end{multline}
with the constraint $\chi(x)\le 0$ for all $x$.
In particular, the left and right parts of the tails
($x<x_0$ and $x>x_0$, respectively) are distributed
independently. The action (\ref{tail-action}) describes
a diffusion with a drift, and the resulting form
of the probability distribution for $\chi(x)$
is approximately normal, with its variance growing
linearly with $x$ and its average decreasing linearly with
$x$, as $x$ moves away from the center $x_0$.

\begin{figure}[tb]
\centerline{\includegraphics[width=0.4\textwidth]{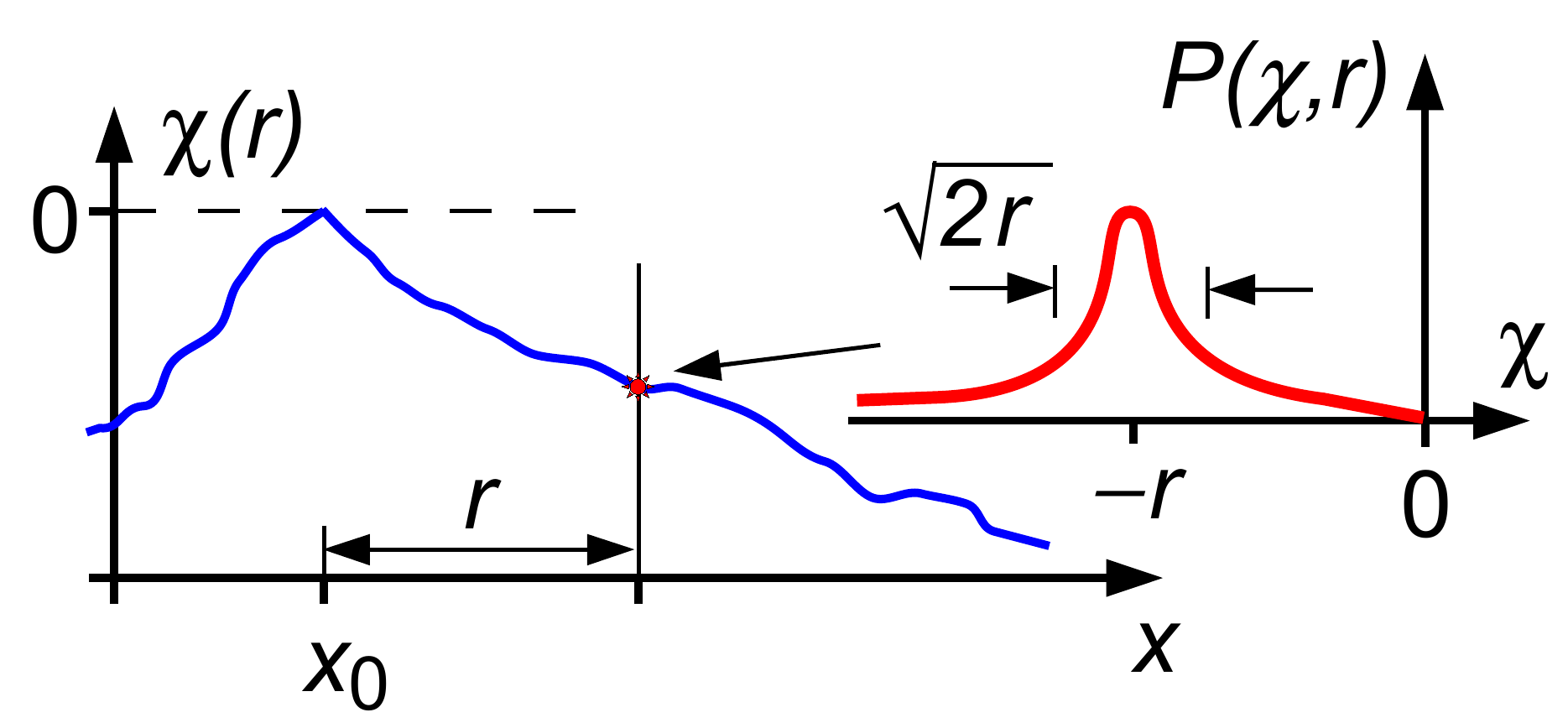}}
\caption{A description of a single-wave-function statistics.
The wave function is described by the position $x_0$ of its
maximum and the distribution of its tails. The one-point
probability distribution $P(\chi,r)$ is assumed to have the
form (\ref{tail-log-normal}) at $r \gg 1$.}
\label{fig:single-wf}
\end{figure}

For our calculations, we will be interested in one-
and multi-point probability distributions of $\chi(x)$
for a fixed position of $x_0$. Let us start with
the one-point probability distribution $P(\chi,r)$,
where $r=|x-x_0|$. The action (\ref{tail-action})
results in the differential equation
\begin{equation}
\frac{\partial P}{\partial r} = \frac{\partial^2 P}{\partial \chi^2}
+ \frac{\partial P}{\partial \chi}
\end{equation}
which describes diffusion with a drift. This equation,
together with the boundary condition $P(\chi>0)=0$, results
in the long-time ($r\gg 1$) asymptotic form of the solution
\begin{equation}
P(\chi,r)=f\left(\frac{\chi}{r}\right) P_0(\chi,r)\, , \qquad \chi<0\, ,
\label{tail-log-normal}
\end{equation}
where
\begin{equation}
P_0(\chi,r)= \frac{1}{2\sqrt{\pi r}}
\exp \left[ - \frac{(\chi+r)^2}{4r} \right]
\label{pure-log-normal}
\end{equation}
is the normal distribution (Fig.~\ref{fig:single-wf}).
The effect of the boundary
condition is the ``cut-off'' factor $f(\chi/r)$.
The exact form of
the function $f(z)$ is determined by a short-time
evolution (at $r\sim 1$) and is therefore beyond our
approximation scheme. The only property that we
assume about $f(z)$ (which becomes useful in
Section \ref{sec:leading}) is its asymptotic
behavior
\begin{equation}
f(z) \propto -z \quad {\rm at} \quad z\to -0\, .
\end{equation}
The normalization of the probability distribution
$P(\chi,r)$ implies $f(-1)=1$. Furthermore,
the {\it typical} values of $\chi(x)$ are
not affected by the cut-off factor, and
we have a ``single-parameter-scaling''
relation
\begin{equation}
  \frac12 \langle (\Delta\chi)^2 \rangle = - \langle \chi \rangle = r\, .
\label{SPS-relation}
\end{equation}
Note that a similar single-parameter scaling is
also well-known for the conductance of long
wires \cite{Abrikosov81,Melnikov81,Kumar85,Shapiro86}.

\begin{figure}
\centerline{\includegraphics[width=0.35\textwidth]{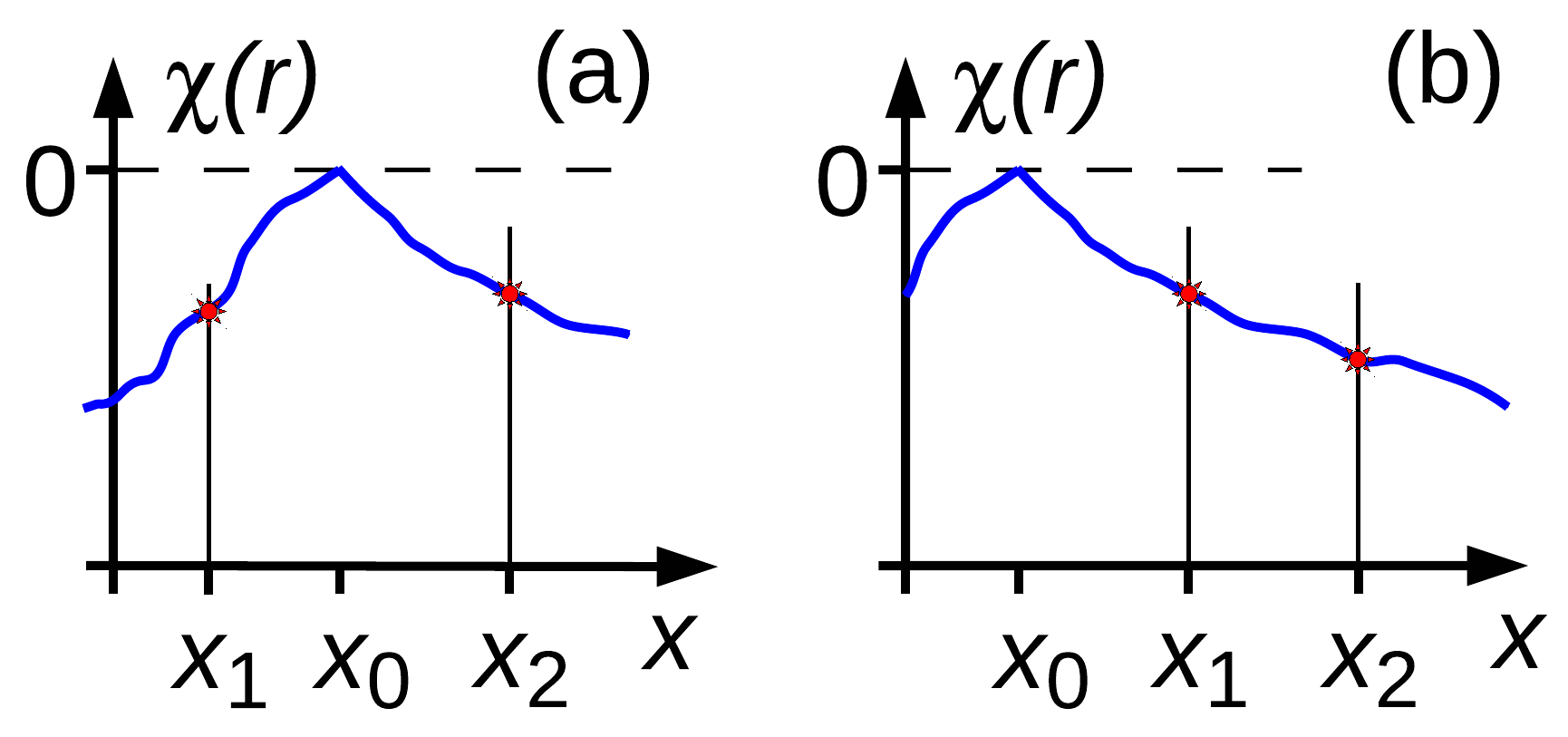}}
\caption{
The two-point probability distribution
$P_{x_0}(\chi_1,x_1;\chi_2,x_2)$
for a single wave function
depends on the ordering of the observation points $x_1$, $x_2$,
and the ``center'' of the wave function $x_0$.
(a) For $x_1$ and $x_2$ belonging to different tails, the distribution
factorizes, see Eq.~(\ref{PP1}).
(b) For $x_1$ and $x_2$ belonging to the same tail, the distribution
is given by Eq.~(\ref{PP2}).
}
\label{fig:single-wf-two-points}
\end{figure}

The above consideration may be directly extended to multi-point
probability distributions. For example, consider the probability
distribution to find $\chi(x_1)=\chi_1$ and $\chi(x_2)=\chi_2$
under the condition that the maximum of $\chi(x)$ is located
at $x_0$ (so that $\chi(x_0)\approx 0$). The form of this
probability distribution $P_{x_0}(\chi_1,x_1;\chi_2,x_2)$
depends on the relative positions of $x_1$, $x_2$, and $x_0$
(see Fig.~\ref{fig:single-wf-two-points}). If $x_1$ and $x_2$
lie on opposite sides of $x_0$ (Fig.~\ref{fig:single-wf-two-points}a),
then the joint probability distribution factorizes:
\begin{equation}
\label{PP1}
  P_{x_0}(\chi_1,x_1;\chi_2,x_2)=P(\chi_1,r_1) P(\chi_2,r_2)\, ,
\end{equation}
where $r_i=|x_i-x_0|$ and $P(\chi,r)$ is given by
Eq.~(\ref{tail-log-normal}). In other words, the
left and right tails are statistically independent.
In the opposite
case, if $x_1$ and $x_2$ belong to the same tail,
the distributions of $\chi(x_1)$ and $\chi(x_2)$ are
correlated.
For the configuration shown in Fig.~\ref{fig:single-wf-two-points}b
(the point $x_1$ lies between $x_0$ and $x_2$), one gets
\begin{equation}
\label{PP2}
  P_{x_0}(\chi_1,x_1;\chi_2,x_2)=
  P(\chi_1,r_1) P_0(\chi_2-\chi_1,r_2-r_1)\, .
\end{equation}
Note that the second factor does not involve the cut-off function $f(z)$,
since, at $-\chi_1 \gg 1$ and $-\chi_2 \gg 1$, the probability
of the functional integral (\ref{tail-action}) to return
to $\chi(x)=0$ is exponentially small.

We may note in passing that the two-point distributions (\ref{PP1})
and (\ref{PP2}) are consistent with the one-point distribution
(\ref{tail-log-normal}). Namely,
\begin{equation}
\int d\chi_1\, P_{x_0}(\chi_1,x_1;\chi_2,x_2) =
 P(\chi_2,|x_2-x_0|)\, ,
\end{equation}
irrespectively of the relative positions of the points
$x_0$, $x_1$, and $x_2$.

Generalization of this construction to many-point distributions
is straightforward. The only requirements are that the distances
between all the points involved exceed the localization length,
$|x_i-x_j|\gg 1$, and that only small tails are considered,
$-\chi_i \gg 1$.

\section{Wave-function hybridization}
\label{sec:hybridization}

It was realized in the early works on localization
that, at $\omega \ll \Delta_\xi$, the
correlation functions (\ref{R-function}) and (\ref{S-function}) may be
understood in terms of the hybridization of two localized states
\cite{Mott,SivanImry1987}.
Following the original argument, we may cut the wire into smaller pieces
and consider
two states $\psi_A$ and $\psi_B$ localized in different pieces
(centered at positions $x_A$ and $x_B$, respectively, and with energies
$E_A$ and $E_B$). As we connect the pieces of the wire together, the
states get hybridized, and such pairs of states give the main contribution
to the correlation functions (\ref{R-function}) and (\ref{S-function}).

\begin{figure}[tb]
\centerline{\includegraphics[width=0.25\textwidth]{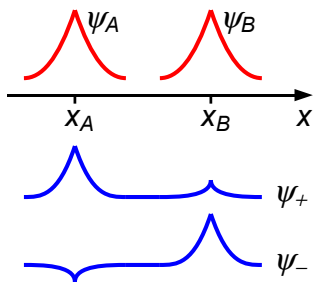}}
\caption{A schematic view of the hybridization of the two localized
wave functions $\psi_A$ and $\psi_B$, as described by Eq.~(\ref{hybrid}).}
\label{fig:hybrid}
\end{figure}

To use this argument at a quantitative level, we need to introduce the
``hybridization'' matrix element $J$ between the states
$\psi_A$ and $\psi_B$. Then the hybridized wave functions are given by
the linear combinations
\begin{equation}
\psi_+ = u_+ \psi_A + u_- \psi_B\, , \qquad
\psi_- = u_-^* \psi_A - u_+^* \psi_B\, ,
\label{hybrid}
\end{equation}
where
\begin{equation}
  |u_\pm|^2 = \frac{1}{2} \left( 1 \mp \frac{\eps}{\Delta} \right)\, ,
\label{u-pm}
\end{equation}
$\eps= E_B- E_A$ and
\begin{equation}
\label{Delta}
  \Delta=\sqrt{\eps^2 + 4|J|^2}
\end{equation}
are the energy splittings
before and after hybridization (Fig.~\ref{fig:hybrid}).
Such a pair of hybridized states contributes
to the correlation functions (\ref{R-function}) and (\ref{S-function}), when
$\Delta= \omega$.

It turns out that this approximate description reproduces quantitatively
many features of the exact results, provided the distribution of the
tails (\ref{tail-log-normal}) is taken into account, and
appropriate assumptions on $J$ are made.

Namely, by analogy with the hybridization of states localized
in potential wells \cite{Landafshits3}, we assume that the hybridization
matrix element $J$ is proportional to the product of the
two envelopes $\widetilde\psi_A(x)$ and $\widetilde\psi_B(x)$:
\begin{equation}
| J | = \Phi\, \widetilde\psi_A(x)\, \widetilde\psi_B(x)\, ,
\label{J-psi-formula}
\end{equation}
where $\Phi$ is a coefficient of order one, which takes into
account the short-range oscillations of the wave functions
$\psi_A(x)$ and $\psi_B(x)$. The distribution of $\Phi$ is
assumed to be statistically independent of the distributions
of the envelopes  $\widetilde\psi_A(x)$ and $\widetilde\psi_B(x)$.
The average  $\langle \Phi \rangle$ is taken to be of order one,
so that Eq.~(\ref{J-psi-formula}) gives the
matrix element $J$ in the units of $\Delta_\xi$.

The specific properties of the distribution of $\Phi$ will be of relevance
for some of our calculations below. In fact, it is this distribution
that distinguishes between the S1D and Q1D geometries and between
the different symmetry classes in the Q1D case. Specifically, in
Section~\ref{sec:subleading}, we will need the behavior of the
probability distribution of $\Phi$ at $\Phi \to 0$. Based on
an analogy with the random-matrix theory, in that part
of the calculation, we will use the following ansatz:
\begin{subequations}
\label{Phi-distributions}
\begin{align}
& dP(\Phi)=\delta(\Phi-\Phi_0)\, d\Phi \quad {\rm with~} \Phi_0\sim 1
\quad {\rm in~S1D}\, ,
\label{Phi-S1D} \\
& dP(\Phi)\propto \Phi\, d\Phi\, , \quad \Phi \to 0 \quad {\rm in~Q1D~unitary}\, ,
\label{Phi-Q1DU} \\
& dP(\Phi)\propto d\Phi\, ,  \quad \Phi \to 0 \quad {\rm in~Q1D~orthogonal}\, .
\label{Phi-Q1DO}
\end{align}
\end{subequations}
Our ansatz for $dP(\Phi)$ in the Q1D-unitary and Q1D-orthogonal cases
can be understood in terms of the sum of the hybridization amplitudes over
a large number of channels. In the case of the unitary symmetry class,
this sum is complex, and therefore its absolute value is distributed
as $\Phi \, d\Phi$ at small $\Phi$, while in the orthogonal symmetry class
it is real with the measure $d\Phi$ at small $\Phi$.

The wave functions  $\widetilde\psi_A(x)$ and $\widetilde\psi_B(x)$
in Eq.~(\ref{J-psi-formula}) are taken at some common point $x$
in the tails of the wave functions. One can verify that, due to
the log-normal statistics of the tails described in
Section~\ref{sec:tails}, the probability distribution of the product
$\widetilde\psi_A(x) \widetilde\psi_B(x)$ is independent of the
specific position of the point $x$. In other words, our ansatz
(\ref{J-psi-formula}) gives a consistent definition of the
probability distribution of $|J|$. Equivalently, one
may also rewrite
\begin{equation}
  |J|=\Phi\, e^{\chi_J/2}\, ,
\label{J-chi}
\end{equation}
where the parameter $\chi_J$ has a distribution of the type
(\ref{tail-log-normal}), with or without a cut-off factor
(depending on the type of the points where the values of
$\widetilde\psi_A(x)$ and $\widetilde\psi_B(x)$ are fixed).

We are now ready to formulate the improved version of the
Mott hybridization argument by combining the three ingredients:
(i) the hybridization of the wave functions (\ref{hybrid}),
(ii) the statistical properties of a single wave function
(\ref{tail-action}), and (iii) the properties of the
hybridization matrix element (\ref{J-psi-formula}).
To obtain the $\omega \ll \Delta_\xi$ limits of the
correlation functions (\ref{R-function}) and (\ref{S-function}),
we restrict the sums over $m$ and $n$ to the two hybridized
states (\ref{hybrid}) and arrive at
\begin{widetext}
\begin{gather}
  R(\omega,|x_1-x_2|)= \int dx_A \, dx_B
  \int d\mu_{x_A} [\chi_A(x)]\,  d\mu_{x_B} [\chi_B(x)]
  \int dP(\Phi)
  \int d\eps\,
  |\psi_+(x_1)|^2 |\psi_-(x_2)|^2\,
  \delta\left(\Delta - \omega\right)\, ,
\label{Mott-R-general}
\\
  S(\omega,|x_1-x_2|)= \int dx_A \, dx_B
  \int d\mu_{x_A} [\chi_A(x)]\,  d\mu_{x_B} [\chi_B(x)]
  \int dP(\Phi)
  \int d\eps\,
  \psi^*_+(x_1)\psi_+(x_2)\psi^*_-(x_2)\psi_-(x_1)\,
  \delta\left(\Delta - \omega\right)\, .
\label{Mott-S-general}
\end{gather}
\end{widetext}
Here the average is taken (i) over the positions of the maxima
$x_A$ and $x_B$ of the envelopes
$\widetilde\psi_A(x)$ and $\widetilde\psi_B(x)$, respectively;
(ii) over the statistical properties of the
wave-function tails $d\mu_{x_A} [\chi_A(x)]$ and
$d\mu_{x_B} [\chi_B(x)]$ defined by Eq.~(\ref{tail-action}),
with the constraint $\chi_\alpha(x)\le 0$ [here we
define $\chi_\alpha(x)=\ln|\widetilde\psi_\alpha(x)|^2$,
as in Eq.~(\ref{chi-def})];
(iii) over the energy difference $\eps$;
and (iv) over the coefficient $\Phi$
in Eq.~(\ref{J-psi-formula}).

The following sections are devoted to extracting the
three different regimes from the general formalism
(\ref{Mott-R-general}) and (\ref{Mott-S-general}):
the behavior at the Mott scale and the leading and
subleading corrections at sub-Mott lengths.

\section{Behavior at the Mott scale}
\label{sec:Mott}

As pointed out in the early works \cite{Mott},
the hybridization of localized states
introduces the logarithmically large ``Mott scale'' (\ref{Mott-length}).
The leading contribution to the behavior of $R(\omega,r)$ at $r\sim L_M$
is obtained by picking out the following term from the general
formula (\ref{Mott-R-general}):
\begin{equation}
  |\psi_+(x_1)|^2 |\psi_-(x_2)|^2
  \longrightarrow
  |u_+|^4 |\psi_A(x_1)|^2 |\psi_B(x_2)|^2 \, .
\label{Mott-pick-R}
\end{equation}
Since the wave functions $\psi_A$ and $\psi_B$ are localized at distances
of order one, and $R(\omega,r)$
at $r\sim L_M$ varies at a
logarithmically larger scale ($\delta r \sim \sqrt{L_M}$, as shown below),
the variables $x_A$ and $x_B$ are nearly pinned to the points $x_1$ and $x_2$,
respectively. Then integration over $x_A$ and $x_B$ yields just
the unit normalization of $\psi_A$ and $\psi_B$, and we get
\begin{equation}
  R_M(\omega,r)
  =
  \int dP(J) \int d\eps \, \delta(\Delta - \omega) |u_+|^4\, ,
\label{Mott-R-1}
\end{equation}
where $\Delta$ and $u_+$ are functions of $\eps$ and $J$ defined by
Eqs.\ (\ref{u-pm}) and (\ref{Delta}). The measure of integration
over $J$ is
\begin{equation}
  dP(J) = f^2 \left( \frac{\chi_J}{r} \right)\,
  P_0(\chi_J,r) \, d\chi_J \, dP(\Phi) ,
\end{equation}
where $J$ is parameterized by Eq.~(\ref{J-chi}),
$P_0(\chi_J,r)$ is the normal distribution (\ref{pure-log-normal}),
and $f(\chi_J/r)$ is the cut-off function
[the same as in Eq.~(\ref{tail-log-normal})].
The integral over $\eps$ may be easily taken, which gives
\begin{equation}
  R_M(\omega,r)=\int dP(J) \, \frac{1}{2} \left(
  \frac{\eps}{\omega} + \frac{\omega}{\eps} \right)\, ,
\label{Mott-R-2}
\end{equation}
where $\eps=\sqrt{\omega^2 - 4 |J|^2}$.

Since the main contribution to the integral comes from logarithmically
large intervals of $\chi_J$, one can approximate
\begin{equation}
\frac{1}{2} \left( \frac{\eps}{\omega} + \frac{\omega}{\eps} \right)
\approx \theta \left( \omega - 2 | J |\right)
\label{trick-1-Mott}
\end{equation}
in Eq.~(\ref{Mott-R-2})
[the step function in the right-hand side takes care of the
integration limits]
and disregard the exact form of the distribution of
$\Phi$. Then, within these approximations, one gets
\begin{multline}
R_M(\omega,r)\approx \int_{-\infty}^{2 \ln \omega}
f^2 \left( \frac{\chi_J}{r} \right)\,
P_0(\chi_J,r)\, d\chi_J \\
=\frac{1}{2} \left[ 1+{\rm erf} \left(\frac{r-L_M}{2\sqrt{r}} \right)\right]\, ,
\label{Mott-R-3}
\end{multline}
i.e., the result reported in the last column of Table \ref{table:R}.
Note that the cut-off function $f(\chi_J/r)$ does not play any role in this
calculation, since $f(-1)=1$ by the normalization of probability.

We can further repeat the same procedure for the correlation function
$ S(\omega,r)$ given by Eq.~(\ref{Mott-S-general})
by selecting the term
\begin{multline}
  \psi^*_+(x_1) \psi_+(x_2) \psi^*_-(x_2) \psi_-(x_1) \\
  \longrightarrow
  -|u_+|^2 |u_-|^2 |\psi_A(x_1)|^2 |\psi_B(x_2)|^2 \, .
\label{Mott-pick-S}
\end{multline}
We then arrive at the formula similar to Eq.~(\ref{Mott-R-1}),
but with $|u_+|^4$ replaced by $-|u_+|^2|u_-|^2$. The formula (\ref{Mott-R-2})
then gets replaced by
\begin{equation}
S_M(\omega,r)=\int dP(J) \, \frac{1}{2} \left(
\frac{\eps}{\omega} - \frac{\omega}{\eps} \right)\, .
\label{Mott-S-2}
\end{equation}
Now we cannot simply replace $\eps$ by $\omega$, but need to expand to
the next order. In fact, we can re-express
\begin{equation}
\frac{1}{2} \left(
\frac{\eps}{\omega} - \frac{\omega}{\eps} \right) =
\frac{\partial}{\partial \chi_J} \frac{\eps}{\omega}
\approx - \delta \left( \chi_J- 2\ln\frac{\omega}{2\Phi} \right) \, ,
\label{trick-derivative}
\end{equation}
which allows us to integrate over $\chi_J$ to obtain
\begin{equation}
S_M(\omega,r) \approx - \frac{1}{2\sqrt{\pi r}}
\exp \left[-\frac{(r-L_M)^2}{4 r} \right]\, ,
\label{Mott-S-3}
\end{equation}
again independently of the distribution of $\Phi$ and therefore
universally valid in S1D and Q1D systems (including the numerical
prefactor \cite{loc-length-warning}).

The result (\ref{Mott-R-3}) has been previously rigorously derived
in S1D and in Q1D-unitary cases \cite{GDP1,IOS2009},
and the result (\ref{Mott-S-3})
in the S1D case \cite{GDP2}. Note that the location and width of the features
in $R_M(\omega,r)$ and $S_M(\omega,r)$
reflect directly the median and the width of the
log-normal distribution for $\chi_J$ in Eq.~(\ref{J-chi}). In our ansatz
(\ref{pure-log-normal}), we take them related to each other,
which corresponds to the single-parameter-scaling
regime \cite{ATAF80,Melnikov81,Kumar85,Shapiro86}.
In Ref.\ \onlinecite{SivanImry1987}, a qualitative behavior of $R(\omega,r)$
at the Mott scale was also explained from the hybridization arguments,
but the correct quantitative expression (\ref{Mott-R-3}) could not
be obtained without taking into account the log-normal distribution
of the wave-function tails.

\section{Leading order at distances much shorter than the Mott scale}
\label{sec:leading}

At distances $1 \ll r \ll L_M$, the correlation functions $R(\omega,r)$ and
$S(\omega,r)$ can be found, to the leading order,
from the general expressions (\ref{Mott-R-general}) and (\ref{Mott-S-general})
if one retains only the contributions from $\psi_A$
in both $\psi_+$ and $\psi_-$:
\begin{multline}
\!\!\!\!\!\!
  |\psi_+(x_1)|^2 |\psi_-(x_2)|^2
  \quad \text{and} \quad
  \psi^*_+(x_1)\psi_+(x_2)\psi^*_-(x_2)\psi_-(x_1)\,
\\
  \longrightarrow
  |u_+|^2 |u_-|^2 |\psi_A(x_1)|^2 |\psi_A(x_2)|^2\, .
\label{short-pick}
\end{multline}
Then, using the relation (\ref{trick-derivative}), we can
integrate over all the variables, except for $x_A$ and $\chi_A$
(in the order $\eps$, $\chi_J$, $x_B$, $\Phi$)
and arrive at the result
\begin{multline}
R(\omega,r) \approx S(\omega,r) \\
\approx 2 \int dx_A\,
d\mu_{x_A}[\chi_A(x)] \,
|\psi_A(0)|^2 |\psi_A(r)|^2 \, .
\label{two-vs-single}
\end{multline}
Thus the short-distance behavior of these correlation functions
is universal for S1D and Q1D models and is only determined by
the single-wave-function statistics. This result was rigorously
derived for S1D
(as follows from
Refs.~\onlinecite{Gogolin}, \onlinecite{GDP1}, and \onlinecite{GDP2})
and Q1D-unitary cases \cite{IOS2009},
and the exact form of this function is known,
\begin{equation}
  R(\omega\to 0,r) =
  4\pi^2 \frac{\partial^2}{\partial r^2}
  \int_0^\infty k \, dk \, \frac{\tanh\pi k}{\cosh^2\pi k}
  e^{-(k^2+1/4)r}\, .
\label{single-function-exact}
\end{equation}

Within our approximate method, we cannot derive this exact expression,
but we can access its $r\gg 1$ limit. In this case, the main
contribution comes from $x_A$ located between $0$ and $r$
(see Fig.~\ref{fig:single-wf-two-points}a),
with the two tails of the wave function $\psi_A$ distributed
independently, according to Eq.~(\ref{PP1}):
\begin{multline}
R(\omega,r) \approx S(\omega,r) \approx \int_0^r dx_A
\int_{-\infty}^0 d\chi_1 \int_{-\infty}^0 d\chi_2
\\ {}
  \times
P(\chi_1, x_A) P(\chi_2, r-x_A) e^{\chi_1 +\chi_2}\, .
\end{multline}
If we use our ansatz (\ref{tail-log-normal}) for $P(\chi,r)$, then
this integral formally diverges at $x_A\to 0$ and $x_A\to r$. This
means that the main contribution to the correlation functions comes
from configurations where the maximum of the wave functions coincides
(within the localization length) with one of the two points. At such
short distances, our ansatz (\ref{tail-log-normal}) is not applicable,
but we can estimate the correlation function, up to a numerical
prefactor, by cutting off the integral in $x_A$ within a localization
length from $0$ and $r$
[i.e., by integrating over $x_A$ in the limits $(\delta,r-\delta)$
with $\delta \sim 1$].
This immediately
leads us to the asymptotic expression (at $r\gg 1$)
\begin{equation}
R(\omega,r) \approx S(\omega,r) \propto r^{-3/2} e^{-r/4}\, ,
\label{single-function-asymptotic}
\end{equation}
where the proportionality coefficient cannot be calculated within our
approximation.

The asymptotic expression (\ref{single-function-asymptotic}) is in
agreement with the exact expression (\ref{single-function-exact})
\cite{Gogolin,Kolokolov1995,Mirlin-review,IOS2009}.
Note that the form of the cut-off in the probability distribution
(\ref{tail-log-normal}) was important for calculating the correct
power in the pre-exponent in Eq.~(\ref{single-function-asymptotic}).
In fact, the correlation functions $R(\omega\to 0,r)$ and $S(\omega\to 0,r)$
are dominated not by ``typical'' localized wave functions,
but by the rare events, when the wave function $\psi_A$ has two peaks
at the positions $0$ and $r$ of comparable height. This can also be seen
from the exponential decay $e^{-r/4}$, which does not describe the
decay of a ``typical'' wave function [whose weight decays as $e^{-r}$,
according to Eq.~(\ref{SPS-relation})],
but is four times slower.

Somewhat similar rare events are important for
the statistics of wave functions in the
metallic limit \cite{AKL}, which also results in log-normal tails. 
However, the metallic regime is beyond the scope of the present paper.

Note that the derivation of the
relation (\ref{two-vs-single}) does not use the condition $r\gg 1$
(which is only needed for calculating its right-hand side), and
is therefore valid for any $r$ (in the limit $\omega \to 0$).
In terms of wave-function correlations,
we may also rewrite Eq.~(\ref{two-vs-single}) as
\begin{multline}
R(\omega\to 0,|x_1-x_2|) = S(\omega\to 0,|x_1-x_2|) \\
= 2 \xi\,  \nu^{-1}
\sum_n \delta(E_n-E) |\psi_n(x_1)|^2 |\psi_n(x_2)|^2\, ,
\label{two-vs-single-1d}
\end{multline}
(in this equation, we restore the physical units).
This relation (without specifying the numerical prefactor)
was already proposed in Ref.\ \onlinecite{SivanImry1987}
based on similar hybridization arguments. However, the approach
used in that work could not correctly reproduce the asymptotic
behavior (\ref{single-function-asymptotic}), since it did not
include the log-normal distribution of the wave-function tails
crucial for such a calculation.

\section{Subleading order at distances much shorter than the Mott scale}
\label{sec:subleading}

\begin{figure}[tb]
\centerline{\includegraphics[width=0.25\textwidth]{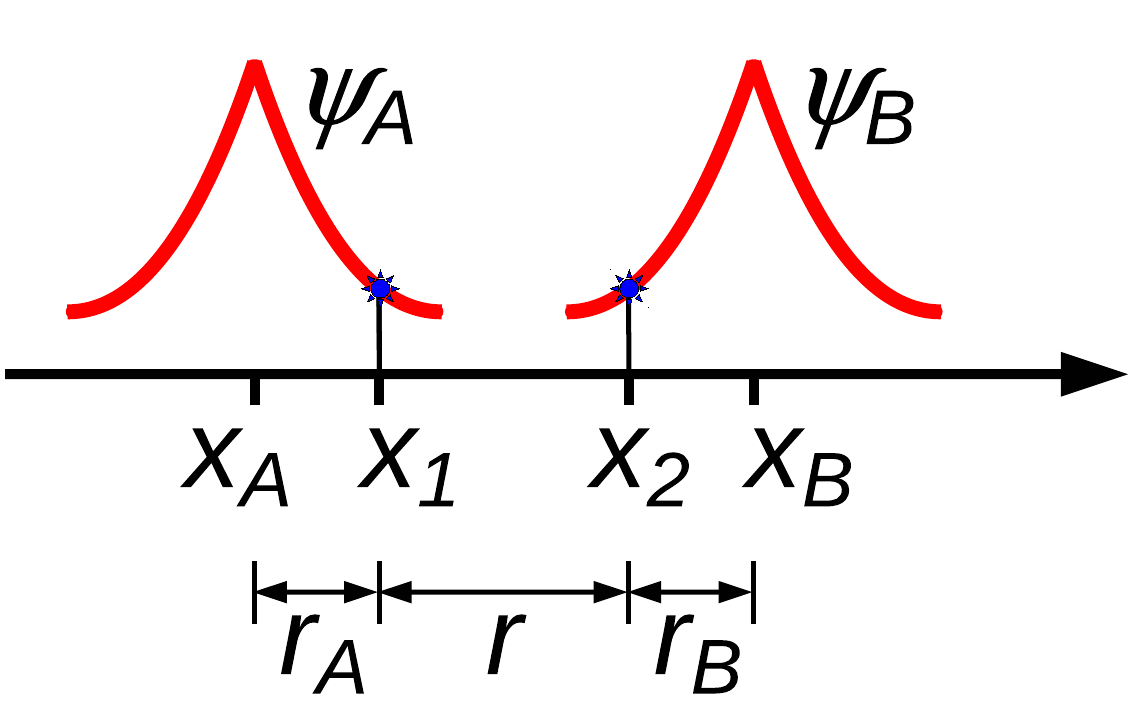}}
\caption{An illustration of notation for the calculation of the subleading
order at $1\ll r \ll L_M$. The centers $x_A$ and $x_B$ of the pair
of localized states are located outside the interval $(x_1,x_2)$. The
variables $r_A$, $r_B$, and $r$ used in Eq.~(\ref{Mott-dR-1})
are the pairwise distances between the four points $x_A$, $x_1$, $x_2$, and $x_B$.}
\label{fig:hybrid-subleading}
\end{figure}

Remarkably, we can extend our method further to finding non-universal
corrections $\delta R(\omega,r)$ and $\delta S(\omega,r)$
[defined in Eqs.~(\ref{deltaRS})]
to the asymptotic behavior (\ref{single-function-asymptotic}).
Such corrections are given by
the same ``cross-terms'' (\ref{Mott-pick-R}) and (\ref{Mott-pick-S}),
as in the calculations of $R_M(\omega,r)$ and $S_M(\omega,r)$ in
Section \ref{sec:Mott}.
One can check that, for this term, the main contribution
comes from configurations with the points
$x_A$ and $x_B$ (the maxima of the wave functions $\psi_A$ and $\psi_B$)
located outside the interval $(x_1,x_2)$, see Fig.~\ref{fig:hybrid-subleading}.

For calculating $\delta R(\omega,r)$,
we start with the general expression (\ref{Mott-R-general}),
where we only keep the term (\ref{Mott-pick-R}). Unlike in
the calculation of Section \ref{sec:Mott}, here the tails
of the wave functions contribute, and therefore we need to
take into account their log-normal distributions.
If we write $|\psi_A(x_1)|^2 = e^{\chi_A}$ and
$|\psi_B(x_2)|^2 = e^{\chi_B}$, then the hybridization matrix element
$|J|$
may be expressed
in the form (\ref{J-chi}) with
\begin{equation}
  \chi_J = \chi_A+\chi_B+\chi .
\label{JAB-2}
\end{equation}
Here $\chi_A$, and $\chi_B$ are distributed with the probability
distribution (\ref{tail-log-normal}) (with the cut-off) and
the distribution of $\chi$ is given by Eq.~(\ref{pure-log-normal})
(without the cut-off)
[compare with an analogous form of Eq.~(\ref{PP2})].
In the calculations
of this section, we further neglect the cutoffs, since all the three parameters
 $-\chi$, $-\chi_A$, and $-\chi_B$ are much larger than one and the integrals
with respect to them can be done at the saddle-point level. The cut-off functions
contribute only to the overall numerical coefficient, which is beyond
the precision of our calculation.

After integrating Eq.~(\ref{Mott-R-general}) over $\eps$, we arrive at
\begin{multline}
  \delta R (\omega,r) \propto \int_0^\infty dr_A \int_0^\infty dr_B
\int_{-\infty}^0 d\chi_A \int_{-\infty}^0 d\chi_B
\\ {}
  \times
P_0(\chi_A, r_A) P_0(\chi_B,r_B)
e^{\chi_A+\chi_B}
\\ {}
  \times
\int d\chi\, P_0(\chi,r)
\int dP(\Phi)\, \frac{1}{2} \left(\frac{\eps}{\omega}+\frac{\omega}{\eps}\right)\, .
\label{Mott-dR-1}
\end{multline}
Like in the calculation
in Section \ref{sec:Mott},
we can use
the approximation (\ref{trick-1-Mott}).
Furthermore, the integrals over $r_A$ and $r_B$ can be calculated
at the saddle-point level (thereby fixing $r_A=-\chi_A$ and
$r_B=-\chi_B$), and afterwards we integrate over $\chi_A$
and $\chi_B$. The resulting expression is
\begin{multline}
\delta R(\omega,r) \propto \int dP(\Phi) \int \frac{d\chi}{2\sqrt{\pi r}}
\exp \left[ - \frac{(\chi+r)^2}{4 r} \right] \\
\times (z+1)e^{-z} \Big|_{z=\max(0, \chi - 2\ln\frac{\omega}{2\Phi})}\, .
\label{Mott-dR-2}
\end{multline}

One can show that, at $r<L_M/3$, the main contribution comes from
the region $\chi - 2\ln\frac{\omega}{2\Phi} >0$. The integral in $\chi$
can be done in the saddle-point approximation, which
sets $\chi=-3 r$. Finally, only the integral over $\Phi$ remains
[at our level of approximation, we also neglect $+1$ in the
second line of Eq.~(\ref{Mott-dR-2})]:
\begin{equation}
\delta R(\omega,r) \propto \omega^2 e^{2r} \int_{\Phi_\omega}^{\sim 1} dP(\Phi)
\, \frac{\ln \Phi - \ln \Phi_\omega}{\Phi^2} \, ,
\label{Mott-dR-3}
\end{equation}
where $\Phi_\omega = \exp \left[ - (L_M - 3 r)/2 \right]$.
Estimating the integral (\ref{Mott-dR-3}) with the distributions
$dP(\Phi)$ given by Eqs.~(\ref{Phi-distributions})
yields the results reported in the middle column of Table~\ref{table:R}.

An analogous calculation can also be performed for $\delta S(\omega,r)$
starting with Eq.~(\ref{Mott-S-general}).
The calculation parallels the one above, with the only
difference that $(1/2)(\eps/\omega + \omega/\eps)$ in Eq.~(\ref{Mott-dR-1})
must be replaced by
\begin{equation}
\frac{1}{2}\left( \frac{\eps}{\omega} - \frac{\omega}{\eps}\right)
\approx
-\delta \left(2\ln \frac{\omega}{2\Phi} - \chi - \chi_A - \chi_B \right)\, .
\end{equation}
This results in
\begin{multline}
\delta S(\omega,r) \propto - \int dP(\Phi) \int \frac{d\chi}{2\sqrt{\pi r}}
\exp \left[ - \frac{(\chi+r)^2}{4 r} \right] \\
\times ze^{-z} \Big|_{z=\max(0, \chi - 2\ln\frac{\omega}{2\Phi})}\, .
\label{Mott-dS-2}
\end{multline}
To the precision of our approximation, this expression is opposite
in sign to Eq.~(\ref{Mott-dR-2}). Therefore, we conclude that,
within our approximation, $\delta S (\omega, r) \approx - \delta R(\omega,r)$.
The corresponding formulas are reported in the middle column of Table \ref{table:S}.
Note that our method does not give the
numerical coefficients in  $\delta R (\omega, r)$ and $\delta S(\omega,r)$,
but predicts that they have the same absolute value and are opposite in sign
[positive for $\delta R(\omega,r)$ and negative for $\delta S(\omega,r)$].

One can compare our results of this section with the exact
calculations. The only case, where a direct comparison is
possible is the Q1D-unitary case, where $\delta R(\omega,r)$
was computed in Ref.~\onlinecite{IOS2009} and, to the
leading order, coincides (up to a numerical prefactor)
with our present result. Note that our results for
$\delta R(\omega,r)$ and $\delta S(\omega,r)$
in the S1D and Q1D-uintary cases are only applicable at $r<L_M/3$.
The new length $L_M/3$ appeared in Ref.~\onlinecite{IOS2009}
as the distance at which the correction $\delta R(\omega,r)$
in the Q1D-uintary case changes its asymptotic
form (in technical terms, there was a switching of the pole and
the saddle in the integral determining the leading form of the correction).

Another situation where an indirect
comparison can be made is the S1D case. There, using
the formalism of Ref.~\onlinecite{GDP1}, one can
show \cite{SO07} that $\delta R(\omega,r)$, at small $r$
starts with the order $\omega^2 \ln\omega$,
i.e. consistent with our result for the S1D case (strictly
speaking, the comparison is not accurate, since our result
is only applicable at $r \gg 1$ while the expansion of
the result from Ref.~\onlinecite{GDP1} is done at $r \ll 1$
but we expect that the leading $\omega$ dependence
is the same in both regimes). In the other cases,
there are no exact calculations to which our
results reported in Tables \ref{table:R} and \ref{table:S}
could be compared, thus they should be considered as conjectures.

\section{Summary, discussion, and outlook}
\label{sec:summary}

To summarize, in this paper we propose a simple technique of
treating
correlation of wave functions in Anderson localization
in terms of hybridization of localized states.
While the essence of our method
repeats the well-known Mott argument, supplementing it
with a log-normal probability distribution for the tails
of localized wave functions (and, consequently,
for the hybridization matrix elements) gives the method
a quantitative power. We have checked that the results
produced with our simplified method reproduce quantitatively
the main features of the available exact results  (obtained
by more sophisticated techniques).

Nevertheless, we should emphasize that the presented method
remains a phenomenological recipe, and its justification
still needs to be completed. The method depends on several
assumptions of various level of rigor. For the benefit of the
reader, we list them below:
\begin{itemize}
\item[1.]
A possibility to define localized wave functions that
are further hybridized into the eigenstates (\ref{hybrid}) close
in energy. These wave functions $\psi_A$ and $\psi_B$ are not
rigorously defined in our argument, and the formalization of this
step would be helpful for a rigorous justification of the method.
\item[2.]
The log-normal distribution of the wave-function tails (\ref{pure-log-normal}),
supplemented by a suitable cutoff (\ref{tail-log-normal}). While we
present some arguments in favor of these formulas, they are not
formally derived. We hope that a rigorous derivation of this step
may be possible with the methods of Ref.~\onlinecite{Mirlin-review}.
\item[3.]
The hybridization matrix elements (\ref{J-chi}) are assumed to be proportional
to $\widetilde\psi_A(x) \widetilde\psi_B(x)$,
the product of the two hybridizing tails,
and therefore to obey the same log-normal distribution.
Since neither $J$ nor $\psi_{A,B}$ are formally defined, this
assumption also remains a phenomenological construction.
\item[4.]
The factor $\Phi$ in (\ref{J-chi}) reflecting the interference of channels.
Its probability distributions (\ref{Phi-distributions})
are introduced phenomenologically.
\end{itemize}

Note that, for different calculations, different assumptions play a role.
At the Mott length scale (Section \ref{sec:Mott}), we only used the
assumptions 1 and 3, with the assumptions 2 and 4 being irrelevant.
For the leading behavior at $1\ll r \ll L_M$ (Section \ref{sec:leading}),
we also used the assumption 2, while in Section \ref{sec:subleading}
we additionally need the assumption 4 to calculate the subleading
terms $\delta R(\omega,r)$ and $\delta S(\omega,r)$.

We wish to remark here that our method was developed under the
specific assumption of a weak disorder (specifically, it is applicable
in the model of the Gaussian white noise), which implies the
single-parameter-scaling relation (\ref{SPS-relation}) between the
variance and the average of the wave-function tail.
However, the method can be extended
to other interesting models of localization by relaxing this assumption.
This would imply a modification of the log-normal probability distributions
(\ref{tail-log-normal}) and (\ref{pure-log-normal})
in the assumptions 2 and 3. One example where
such a modification may be applicable is localization far below the
mobility edge (see, e.g., Refs.\ \onlinecite{HSW80,KLP03,Gruzberg}).
Another interesting example is the exactly solvable localization
problem with the Cauchy-distributed disorder considered in
Ref.~\onlinecite{DLA00}. In that model, the
single-parameter-scaling relation (\ref{SPS-relation}) does apply,
but with a different coefficient. We believe that such a system
can also be treated by our method with a suitably modified
probability distributions
(\ref{tail-log-normal}) and (\ref{pure-log-normal}).

Finally, it would be interesting to extend our approach to
higher dimensions. The key issue for such an extension is the
log-normal distribution of the wave-function tails (our assumption
2 above). While we are not aware of such results for wave functions
in higher-dimensions, a similar claim was made for the probability
distribution of the conductance \cite{Shapiro86}. Namely, it
was shown that,
in the weak-scattering case and in the insulating
phase, the conductance distribution in
the large-system-size limit is log-normal with the
single-parameter-scaling relation between the average and the variance
of the form (\ref{SPS-relation}).
Therefore one may assume that
a similar universal distribution is also valid for the wave-function
tails. If it is indeed the case, then our calculations in Sections
\ref{sec:Mott} and \ref{sec:leading} can be straightforwardly
extended to higher dimensions. In particular, the counterpart of
Eq.~(\ref{two-vs-single-1d}) in any dimension would read
\begin{multline}
R(\omega\to 0,|x_1-x_2|) = S(\omega\to 0,|x_1-x_2|) \\
= S_{d-1} L_M^{d-1} \xi\,  \nu^{-1}
\sum_n \delta(E_n-E) |\psi_n(x_1)|^2 |\psi_n(x_2)|^2\, ,
\label{two-vs-single-general}
\end{multline}
where $S_{d-1}$ is the area of the $(d-1)$-dimensional sphere
(e.g., in our one-dimensional case, $S_0=2$),
and the definition (\ref{Mott-length})
of $L_M$ remains the same in any dimension.
Also, under the same conditions, the behavior of $R_M(\omega,r)$ and
$S_M(\omega,r)$
(the right column of Tables \ref{table:R} and \ref{table:S}
calculated in Section \ref{sec:Mott}) would be universal in any dimension.
This implies, in particular, the validity of the Mott formula for the
frequency-dependent conductivity $\sigma(\omega)\propto\omega^2 (\ln\omega)^{d+1}$
in any dimension (which can be deduced from $S_M(\omega,r)$, see, e.g.,
Ref.\ \onlinecite{GDP2}), in agreement with the original
argument \cite{Mott}.
\null

\begin{acknowledgments}
We thank A.~D.~Mirlin for drawing our attention to the role
of log-normal distribution of wave-function tails
in Mott's phenomenology and M.~V.\ Feigelman for comments
on the manuscript.
This work was partially supported by the RFBR grants No.\ 10-02-01180
and No.\ 11-02-00077, the program ``Quantum physics of
condensed matter'' of the RAS, the Dynasty Foundation,
and the Russian Federal Agency of Education (contract No.\ P799).
M.~A.~S., P.~M.~O., and Ya.~V.~F.\ thank ITP, EPFL for
hospitality.
\end{acknowledgments}


\begin{thebibliography}{99}
\bibitem{loc-length-definition}
In Q1D wires, the localization length can be microscopically
expressed as \cite{EL83,Efetov-book}
$\xi=\pi\beta\nu A D$, where $\nu$ is the
density of states, $A$ is the cross section of the wire,
$D$ is the diffusion coefficient, and
$\beta$ is the level-repulsion parameter ($\beta=1$ in
the Q1D-orthogonal case and $\beta=2$ in the Q1D-unitary case).
We define the localization energy spacing
in all cases as $\Delta_\xi=(2\pi\nu A \xi)^{-1}$.
\bibitem{loc-length-warning}
Note that since $\beta$ enters the
expression for the localization length $\xi$ \cite{loc-length-definition},
the ``universality'' of some of our formulas only means that they
have the same form when expressed in terms of the corresponding $\xi$
and not in any case invariance with respect to time-reversal-symmetry-breaking
perturbations.
\end{thebibliography}
\end{document}